# An AI-powered Public Health Automated Kiosk System for Personalized Care: An Experimental Pilot Study

Sonya Falahati[1,2], Morteza Alizadeh[1,3], Fatemeh Ghazipour[1,4], Zhino Safahi[1,5], Navid Khaledian[6], Mohammad R. Salmanpour[1,7*]

[1] Technological Virtual Collaboration (TECVICO Corp.), Vancouver, BC, Canada
[2] Electrical and Computer Engineering Department, Nooshirvani University of Technology, Babol, Iran
[3] Department of Mathematics, University of Isfahan, Isfahan, Iran
[4] CDI College, Burnaby, BC, Canada
[5] Social & Biological Network Laboratory, Department of Computer Engineering, University of Kurdistan, Sanandaj, Iran
[6] Interdisciplinary Centre for Security, Reliability and Trust (SnT), University of Luxembourg, Esch-sur-Alzette, Luxembourg
[7] Department of Integrative Oncology, BC Cancer Research Institute, Vancouver, Canada

(*) **Corresponding Author:** Mohammad R. Salmanpour, Ph.D. (msalman@bccrc.ca)

**ABSTRACT**

**Background:** The HERMES Kiosk (Healthcare Enhanced Recommendations through Artificial Intelligence (AI) & Expertise System) is designed to provide personalized Over-the-Counter (OTC) medication recommendations, addressing the limitations of traditional health kiosks. It integrates an advanced GAMENet model enhanced with Graph Attention Networks (GAT) and Multi-Head Cross-Attention (MHCA) while ensuring user privacy through federated learning. This paper outlines the conceptual design and architecture of HERMES, with a focus on deployment in high-traffic public areas.

**Methods**: HERMES analyzes self-reported symptoms and anonymized medical histories using AI algorithms to generate context-aware OTC medication recommendations. The system was initially trained using Electronic Health Records (EHR) from the MIMIC-III dataset (6,350 patients) and Drug-Drug Interaction (DDI) data from the TWOSIDES database, incorporating the top 90 severity DDI types. Real-time DDI checks and Anatomical Therapeutic Chemical (ATC)-mapped drug codes further improve safety. The kiosk is designed for accessibility, offering multilingual support, large fonts, voice commands, and Braille compatibility. A built-in health education library promotes preventive care and health literacy. A survey was conducted among 11 medical professionals to evaluate its potential applications in medicine.

**Results:** Preliminary results show that the enhanced GAMENet model achieved a Precision-Recall Area Under the Curve (PRAUC) of 0.74, outperforming the original model. These findings suggest a strong potential for delivering accurate and secure healthcare recommendations in public settings.

**Conclusion:** HERMES demonstrates how AI-driven, privacy-preserving kiosks can enhance public health access, empower users, and alleviate burdens on healthcare systems. Future work will focus on real-world deployment, usability testing, and scalability for broader adoption.

**Keywords**: Public Health Kiosk, AI and Federated Learning, Personalized Healthcare, Over-the-Counter (OTC) Medication Recommendation, Preventive Care

## 1. INTRODUCTION

In an era of increasing healthcare demands and limited resources, providing timely and personalized medical care remains a significant challenge [1]. The rise of AI-powered technologies offer new opportunities to address these issues, particularly through the deployment of health kiosks in public spaces. These kiosks have the potential to bridge gaps in healthcare access by offering convenient, cost-effective, and personalized services to a wide range of users [2] [3]. Traditional health kiosks, while beneficial, often fall short of meeting the growing demand for comprehensive and personalized care. They typically offer only basic services, such as blood pressure checks, and lack the ability to provide tailored recommendations based on individual health profiles [4]. Additionally, concerns about data security, privacy, and limited human interaction have hindered their widespread adoption [5], these limitations highlight the need for more advanced solutions that leverage cutting-edge technologies to deliver accurate, secure, and user-friendly healthcare services.

The optimal location for health kiosks remains a topic of debate, with varying levels of acceptance depending on whether they are placed in public or medical settings [6] [7]. In high-traffic public spaces, traditional kiosks often struggle to meet user needs due to language barriers and limited information availability [8]. These challenges underscore the importance of designing kiosks that are not only technologically advanced but also inclusive and adaptable to diverse environments.

Recent advancements in AI and healthcare technology have paved the way for innovative solutions like HERMES. Previous studies [9] [10] [11] have demonstrated the effectiveness of AI-powered recommender systems in providing personalized medication recommendations. Similarly, health kiosks have shown promise in expanding access to basic healthcare services [12] [13]. However, these systems often lack the sophistication needed to address complex healthcare needs, such as real-time drug interaction analysis and personalized care based on comprehensive health data [14]. HERMES builds on these foundations by integrating state-of-the-art AI models and prioritizing user privacy and accessibility [15].

The integration of advanced AI models, such as GAMENet [16], into public health kiosks represents a significant step in knowledge translation and knowledge to action [17]. By leveraging patient history and DDI data, HERMES aims to bridge the gap between cutting-edge research and routine clinical practice. This approach not only enhances the accuracy of Over-the-Counter (OTC) medication recommendations [18] but also ensures that the latest advancements in AI are effectively translated into actionable healthcare solutions for the public [19].

Existing health kiosks primarily offer basic healthcare services, such as measuring vital signs (e.g., blood pressure, heart rate, and Body Mass Index (BMI)), providing general health information, and facilitating appointment scheduling. While some advanced kiosks include symptom-checking features and telemedicine consultations, they often operate in isolation, lacking the capability to integrate comprehensive Electronic Health Records (EHR) or perform real-time drug interaction analysis [20] Furthermore, current systems do not incorporate AI-driven models for personalized medication recommendations, limiting their effectiveness in supporting complex clinical decision-making [21]. To address these limitations, HERMES enhances public health kiosks by integrating advanced AI-powered drug recommendation algorithms, leveraging patient history and DDI data to ensure context-aware and clinically safe medication guidance [22]. This approach enables a seamless transition from cutting-edge AI research to practical healthcare solutions, improving patient safety and accessibility in public health settings.

Traditional health kiosks often fall short by offering only basic health checks and lacking comprehensive care. The HERMES kiosk addresses this gap by leveraging advanced AI to provide personalized OTC medication recommendations in accessible public locations. Moreover, traditional kiosks often fail to account for individual differences such as age, medical history, allergies, and potential medication interactions, leading to generic and sometimes ineffective advice [23] [24] [20]. HERMES addresses these shortcomings by offering a more advanced and inclusive approach to self-care. By leveraging deep learning, HERMES provides personalized medication recommendations based on a detailed analysis of user-reported symptoms, considering individual factors such as allergies and medical history. This level of personalization marks a significant departure from the generic advice provided by traditional kiosks [25]. Previous studies [26] [27] have highlighted the importance of knowledge translation in healthcare, particularly in the context of AI-driven solutions. This section explores existing research on healthcare kiosks and medication recommendation systems, highlighting their potential and laying the groundwork for the HERMES kiosk's development.

Healthcare kiosks are rapidly transforming healthcare delivery by offering a range of potential benefits that enhance efficiency, accessibility, and patient experience. These self-service kiosks have become vital to the digital transformation of healthcare, providing advantages for both patients and providers [28]. They streamline workflows by automating routine tasks such as check-in, appointment scheduling, and insurance verification, freeing up valuable staff time for more complex patient interactions. This efficiency translates into a more positive patient experience, as kiosks offer a convenient and potentially faster way to complete administrative tasks, access information, and navigate healthcare settings. Research on check-in kiosks, for instance, suggests that they can lead to increased patient satisfaction [29].

Moreover, kiosks expand access to essential healthcare services in geographically remote areas or communities with limited resources. By providing self-service options for consultations, basic screenings, and health information, kiosks bridge healthcare gaps and empower patients to take a more active role in their health management. Some kiosks even offer access to remote consultations, further enhancing their utility [30]. Additionally, kiosks improve communication through features such as digitized check-in and registration processes, as well as enhanced security with biometric authentication protocols [31] [32]. The integration of AI has opened doors to expanded functionalities, such as virtual triage systems and intelligent analysis, which can improve the prioritization of care. For example, AI-powered kiosks have been used to assess symptoms and recommend appropriate care pathways, reducing the burden on healthcare providers [33]. Advances in robotic processing automation have also streamlined workflows by reducing the burden of time-consuming and repetitive administrative tasks on healthcare staff [34].

Despite these advancements, traditional kiosks often lack the ability to provide comprehensive, personalized care [31]. They typically focus on basic diagnostics and administrative tasks, leaving a gap for more advanced solutions like HERMES, which leverages AI to deliver personalized medication recommendations and real-time safety checks. Furthermore, challenges such as data privacy concerns, limited user interaction, and the need for continuous maintenance have hindered the widespread adoption of traditional kiosks [35] [36]. These limitations highlight the need for innovative solutions that address both technological and user-centric challenges.

Recent studies have also explored the use of kiosks in specialized healthcare settings, such as mental health screening and chronic disease management. For instance, kiosks equipped with AI-driven Chabots have been used to provide mental health support and triage patients to appropriate care providers [37]. Similarly, kiosks designed for diabetes management have demonstrated the potential to improve patient outcomes by providing personalized feedback and educational resources [38]. AI-driven predictive models have also played a significant role in improving patient stratification and treatment planning for chronic conditions such as lung cancer, as demonstrated in recent studies [39] [40] [41] [42] [43] [44] [45]. These applications highlight the versatility of kiosks in meeting varied healthcare needs. While existing systems have improved access and efficiency, HERMES advances this further by leveraging AI and user-centric design to provide personalized, comprehensive care.

Medication recommendation systems (MRS) are revolutionizing healthcare by leveraging AI and HER data to create personalized treatment plans, ultimately optimizing patient outcomes [46]. Early MRS relied on predefined rules, laying the groundwork for more sophisticated techniques like content-based filtering, which recommends medications similar to past successes, and collaborative filtering, which identifies patterns in treatment responses among patients with similar characteristics [47] [48] [49]. Recent advancements in hybrid machine learning approaches, particularly in predicting drug dosage and optimizing treatment strategies for neurological disorders, have shown promising results [50]. These developments enhance the adaptability of MRS in handling complex and personalized treatment regimens The field further diversified with the introduction of hybrid methods (combining these approaches), demographic filtering (categorizing users based on general characteristics), and utility-based methods (recommending items based on overall value) [51].

Matrix factorization has become a cornerstone of collaborative filtering, while deep learning has provided powerful tools to capture complex patterns for highly personalized recommendations [52] [53]. Deep learning approaches like REverse Time Attention (RETAIN) and Dual adaptive sequential network (DASNet) exemplify this progress. RETAIN employs attention mechanisms to identify crucial information within a patient's medical history, while DASNet analyzes EHR data through deep learning [54] [55]. Despite these advancements, challenges remain. Incorporating external knowledge (e.g., DDI) and ensuring fair recommendations across diverse populations pose hurdles [56]. Additionally, the "black box" nature of deep learning models can hinder physician trust due to a lack of interpretability [57].

To manage the computational demands of real-time DDI checks and personalized recommendations, HERMES leverages optimized neural network architectures and edge computing techniques. These innovations ensure that even complex models like Graph Neural Networks (GNNs) can operate efficiently within the constraints of a public kiosk setting without compromising the speed or accuracy of recommendations. GNNs offer a promising alternative by capitalizing on the complex relationships within the healthcare ecosystem, encompassing patients, medications, diagnoses, and other entities [58] [59]. Recent applications [50] [60] in predicting drug responses for Parkinson's disease and optimizing medication plans have further demonstrated the potential of AI in personalized treatment planning. Other studies [61] [62] such as A-GSTCN and DRMP exemplify this approach, utilizing GNNs and message propagation networks to analyze EHR data, consider medication history, and integrate DDI. However, limitations persist in fully capturing intricate relationships and incorporating external knowledge.

Current MRS face challenges like data scarcity, fragmentation, and the inability to consider patient-specific factors fully. Traditional methods' reliance on correlations and underutilizing rich semantic knowledge further limit their effectiveness [63] [64]. Additionally, side-effect prediction methods can be computationally expensive and inaccurate, hindering adoption [65]. HERMES enhances patient access and trust by integrating federated learning for privacy and real-time DDI checks for safety. Its user-friendly interface, personalized recommendations, multilingual support, and accessibility features like large fonts and voice commands address diverse user needs. This synergy between advanced AI and user-centered design promises to make personalized medicine more accessible and equitable across various demographics.

Knowledge transfer, the process of translating research findings into practical applications, plays a pivotal role in bridging the gap between scientific advancements and real-world healthcare solutions. In the context of AI-driven healthcare systems, knowledge transfer ensures that cutting-edge technologies, such as advanced machine learning models, are effectively integrated into routine clinical practice to improve patient outcomes and address public health challenges [66].This paper introduces the HERMES kiosk, an automated system that exemplifies knowledge transfer by leveraging state-of-the-art AI models, such as the enhanced GAMENet architecture, to deliver personalized OTC medication recommendations. By embedding these models into a user-friendly kiosk designed for high-traffic public locations, HERMES translates complex AI research into actionable healthcare solutions, making advanced medical recommendations accessible to diverse populations. This approach aligns with the principles of knowledge translation, which emphasize the importance of adapting research innovations to meet real-world healthcare needs [67]. Previous studies [68] [69] [70] have highlighted the role of knowledge transfer in improving healthcare delivery, particularly through the integration of AI technologies into clinical workflows. For example [69], discuss how innovations, including AI-powered systems, can enhance decision-making in healthcare, while [70] emphasize the importance of user-centered design in ensuring the adoption of such systems.

Recent studies have underscored the significance of knowledge transfer in healthcare, particularly concerning the integration of AI technologies into clinical workflows [71] [72]. One scoping review highlighted the scarcity of comprehensive frameworks guiding AI implementation in healthcare, suggesting that existing models may not fully address the unique challenges posed by AI technologies [73] [74]. Another study [75] demonstrated that a transdisciplinary digital health curriculum could effectively enhance digital competencies among healthcare professionals, facilitating the adoption of innovative technologies. By combining advanced AI with a focus on accessibility and inclusivity, HERMES demonstrates how knowledge transfer can address critical gaps in public health access and empower individuals to take control of their health. To validate HERMES' alignment with clinical needs and knowledge transfer goals, we conducted a structured survey of 11 healthcare professionals, assessing usability, safety, and ethical concerns. Their insights informed iterative refinements to ensure the kiosk's recommendations integrate seamlessly into real-world workflows while addressing clinician trust gaps

## 2. MATERIALS AND METHODS

*2.1. Data*

We utilized EHRs from MIMIC-III, a publicly accessible critical care database containing de-identified clinical data from over 40,000 ICU patients at Beth Israel Deaconess Medical Center (2001–2012) [76]. From this dataset, we selected a cohort of 6,350 patients with multiple hospital visits to enhance longitudinal analysis. Focusing on critical early interventions, we analyzed medications administered within the first 24 hours of admission, as this period significantly influences patient outcomes. The dataset includes 15,016 clinical events, with an average of 2.36 visits per patient, 10.51 diagnoses, 3.84 procedures, and 8.80 medications per visit. Additionally, we integrated DDI knowledge from the TWOSIDES dataset [77] incorporating 90 high-severity DDI types—more extensively than used in previous studies. To ensure compatibility with MIMIC-III, drug codes were transformed from the National Drug Code (NDC) system to the ATC Third Level classification.

*2.2 Model Architecture*

Our proposed architecture, illustrated in Figure 1, extends the GAMENet framework [16] by incorporating two key enhancements: GAT layers for DDI graph encoding that learns interaction-specific attention weights, and MHCA mechanism that dynamically fuses patient representations with global and dynamic drug memories [78]. This dual-attention approach enables more nuanced modeling of complex medication relationships while maintaining clinical interpretability. Our modifications aim to improve the model's ability to capture complex relationships in the DDI graph and to better integrate patient history with global and dynamic drug memories.

GAT for DDI Graph. We replaced the GCN layer used for encoding the DDI graph with a GAT layer. The GAT layer assigns different attention weights to different drug interactions, allowing the model to focus on more critical interactions and better handle the non-transitive nature of DDI graphs. The GAT layer was implemented using the GATConv module from the PyTorch Geometric library, with two attention heads and a dropout rate of 0.5. We added a MHCA layer before the concatenation of the patient representation (query), global drug memory (fact1), and dynamic drug memory (fact2). The MHCA layer allows the model to dynamically weigh the importance of different parts of the input, improving the integration of patient history, global drug memory, and dynamic drug memory. The MHCA layer was implemented using MHCA module from the PyTorch library, with two attention heads. The modified GAMENet model was implemented in PyTorch. The DDI adjacency matrix was converted to an edge index format for the GAT layer. The MHCA layer was applied to the concatenated input of query, fact1, and fact2 before passing it to the output layer. The model was trained using the Adam optimizer with a learning rate of 0.001 and a batch size of 32. The training process included a dropout rate of 0.5 to prevent overfitting.

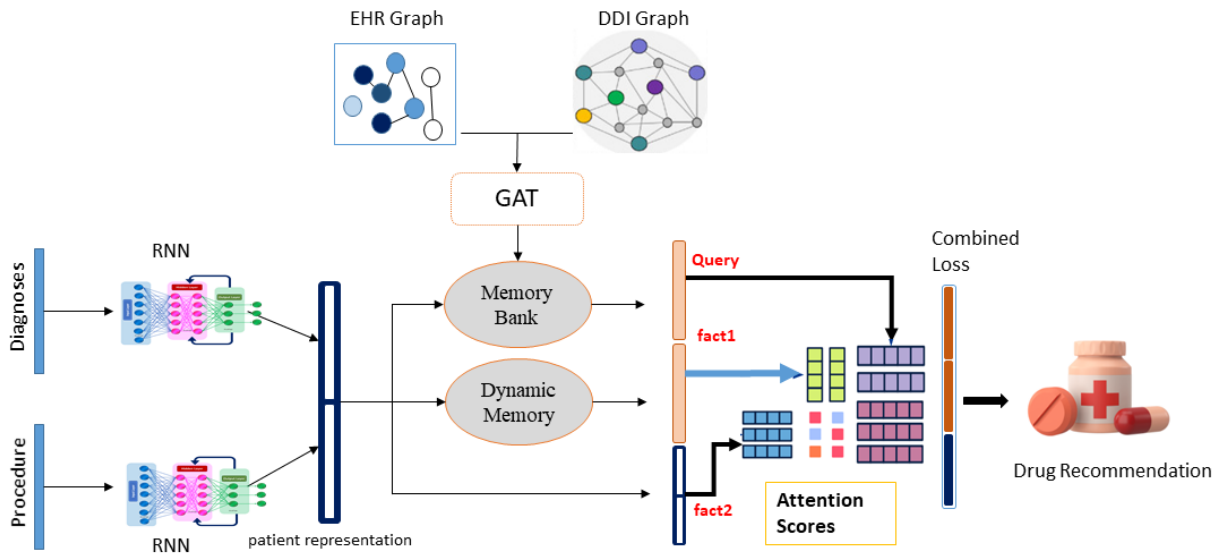

**Fig. 1.** Architecture of the enhanced medication recommendation system showing: (left) EHR processing through diagnosis RNNs, (center) DDI graph encoding via GAT layers, and (right) memory fusion through MHCA. The attention mechanisms (orange) enable dynamic weighting of patient history and drug interactions for clinically-relevant recommendations.

*2.3 AI Engine and Decision Algorithm*

The core functionality of the HERMES kiosk is driven by an AI engine designed to provide personalized OTC medication recommendations. As shown in Figure 2, this engine is trained on a carefully curated dataset with detailed symptom descriptions, corresponding medications, and potential diagnoses. The system adheres to stringent ethical and legal guidelines to ensure data privacy, utilizing anonymized repositories like MIMIC-III for model training. In our implementation, we enhance the GAMENet architecture by introducing a MHCA mechanism before concatenating the queries. This modification allows the model to better capture complex relationships between patient health data, medication history, and DDIs, leading to more accurate and personalized recommendations.

The GNN component of the AI engine analyzes DDIs using a multi-layer perceptron with message-passing operations. This allows the system to move beyond primary symptom-medication associations, identifying complex symptom patterns indicative of specific conditions. For instance, recognizing migraine symptoms enables HERMES to recommend targeted medications beyond generic pain relievers [79] [80] [81]. The AI engine uses DDI data from the TWOSIDES dataset to enhance user safety and evaluate potential interactions based on user-reported medications, advising users accordingly. HERMES prioritizes user experience through a user-friendly interface that supports precise symptom categorization, visual aids, and voice recognition. Moreover, the AI algorithms were evaluated using well-established performance metrics, including Jaccard index, F1 score, and PRAUC [82]. The system provides detailed medication information, including potential side effects, and directs users to emergency services or healthcare providers when necessary. The kiosk also advises seeking professional medical attention for persistent symptoms or those beyond its scope.

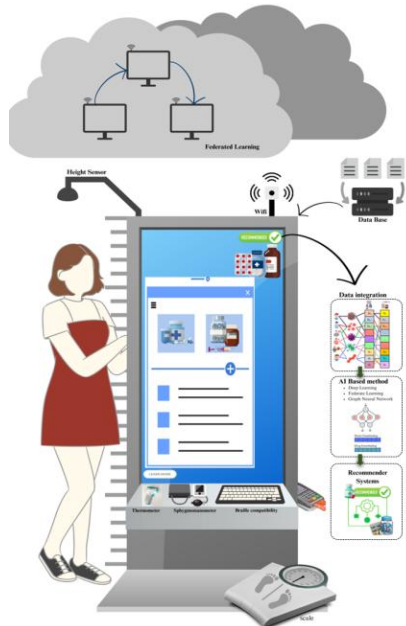

**Fig.2.** HERMES Kiosk Architecture. The HERMES kiosk architecture integrates advanced AI models (GAT and MHCA) for personalized OTC medication recommendations, leveraging federated learning for privacy and real-time DDI checks for safety. Designed for high-traffic public spaces, it includes multilingual support, accessibility features, and a health education library to enhance public health access and literacy.

## 2.4 User Experience and Pilot Study

The HERMES kiosk's development follows a structured five-phase framework, Displayed in Figure 3, prioritizing accessibility and real-world applicability:

*Step 1: Usability Testing with Diverse User Panels.* Usability testing will be conducted with user panels representing diverse demographics, including age groups, ethnic backgrounds, and varying levels of technological proficiency. This step ensures the system is intuitive and accessible, particularly for individuals with limited digital literacy or disabilities. Data will be collected through interviews, questionnaires, and observation to identify areas for improvement.

*Step 2: Iterative Design and Optimization.*
Based on feedback from usability testing, iterative design cycles will be implemented to refine the kiosk's interface and interaction flow. For example, simplifying symptom input or expanding multilingual support. This ensures the system remains user-friendly and adaptable to diverse user needs.

*Step 3: Pilot Deployment in High-Traffic Locations.* The kiosk will be deployed in high-traffic public locations such as transportation hubs, hospitals, pharmacies, and community centers. Collaboration with local health authorities will ensure compliance with regulations and integration with existing healthcare initiatives. This step evaluates the kiosk's real-world impact on healthcare access and medication adherence.

*Step 4: Data Collection and Performance Evaluation.* Quantitative data on session duration, frequency of use, and accuracy of symptom-medication matches will be collected. Qualitative insights from user interviews and focus groups will provide feedback on user satisfaction and areas for improvement. This step assesses the kiosk's effectiveness in diverse environments.

*Step 5: Scalability and Comparative Analysis.* The kiosk's scalability will be tested in locations with varying network reliability and hardware constraints. A comparative analysis will benchmark the HERMES kiosk against traditional healthcare delivery methods, evaluating accuracy, computation efficiency, and user satisfaction. This step ensures the system is ready for large-scale deployment.

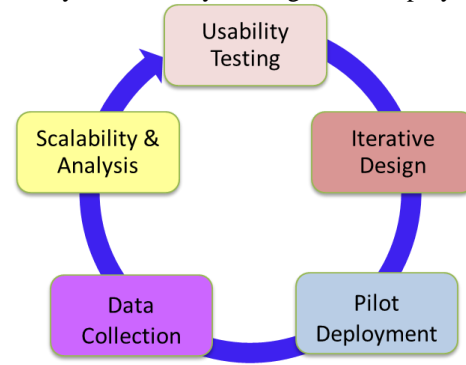

**Fig.3.** The diagram illustrates the 5-phase implementation methodology: (1) Usability Testing with diverse demographic groups, (2) Iterative Design cycles for interface optimization, (3) Pilot Deployment in high-traffic public locations, (4) Performance Evaluation through quantitative and qualitative data collection, and (5) Scalability Analysis comparing system performance against traditional healthcare delivery methods.

## 2.5 Survey and Knowledge Transfer Analysis

To guide the responsible development and deployment of the HERMES Kiosk, a 27-question survey was designed and distributed to 11 healthcare professionals from various specialties, including general practitioners, psychiatrists, physicians, radiologists, and medical students, as seen in Table 2. The goal was to gather expert insights on the clinical relevance, safety, usability, and ethical implications of AI-powered drug recommendation kiosks. The questions were grouped into seven thematic sections for focused exploration. Section 1 (Q1–Q4) covered Demographics and Professional Background, including participants' medical specialty (Q1), years of clinical experience (Q2), prior exposure to AI in healthcare (Q3), and familiarity with AI-driven drug recommendation systems (Q4). These questions were intended to contextualize responses based on the participants' expertise and technological awareness, which are known to influence attitudes toward AI adoption.

Section 2 (Q5–Q9) focused on Usability and Perceived Effectiveness, asking participants to rate the usefulness of AI kiosks in public settings (Q5), identify key benefits (Q6), and assess their effectiveness in empowering patients to manage minor conditions (Q7), reducing unnecessary doctor visits (Q8), and improving medication adherence (Q9). These questions were selected to understand the practical value of the kiosk in everyday healthcare scenarios. Section 3 (Q10–Q12) addressed Accuracy and Clinical Reliability, exploring perceptions of how accurately kiosks can provide appropriate OTC medication suggestions (Q10), the expected frequency of clinically sound recommendations (Q11), and the system's ability to identify safe alternatives for patients with allergies or contraindications (Q12). These insights are essential for evaluating the reliability and clinical robustness of the recommendation engine.

Section 4 (Q13–Q15) examined Clinical Safety and Risk Concerns, addressing participants' primary safety concerns (Q13), the role of real-time DDI checks (Q14), and whether the system should avoid recommending medications to high-risk patients (Q15). These questions helped assess safety guardrails and risk management strategies. Section 5 (Q16–Q22) focused on Implementation and Scalability, including the value of referral features to hospitals or telehealth services in urgent cases (Q16–Q17), the importance of communicating side effects (Q18–Q19), suitable deployment settings (Q20), desired additional features (Q21), and perceived challenges in scaling across health systems (Q22). Section 6 (Q23–Q25) compared the kiosk's effectiveness to traditional care, asking whether it could match or replace human pharmacists (Q23, Q25) and whether it would be more trusted than online symptom checkers (Q24). Finally, Section 7 (Q26–Q27) explored Ethical and Privacy Concerns, including the need for user consent before data collection (Q26) and the importance of regulatory oversight by health authorities like the FDA or WHO (Q27). The survey questions, developed with the expertise of a clinical knowledge translation specialist, were subsequently validated by healthcare professionals. These questions ensured the design considers both patient rights and institutional compliance. Together, the survey responses and open-ended feedback provided foundational insights for aligning the HERMES Kiosk with clinical expectations, patient safety, and ethical AI integration in healthcare.

## 3. RESULTS

### 3.1. Experimental analysis

As part of the initial evaluation, we conducted preliminary tests to assess the performance of the enhanced GAMENet architecture with GAT and MHCA mechanisms. To evaluate HERMES, we used key metrics including the Jaccard Similarity Score to measure the overlap between predicted and actual medications, the Average F1 Score to balance precision and recall for accuracy, and PRAUC to assess the model's ability to identify relevant medications in imbalanced datasets. Additionally, the DDI Rate was used to measure the percentage of recommended medication pairs that result in DDIs, ensuring safer recommendations. These metrics collectively provide a comprehensive evaluation of HERMES, balancing accuracy, safety, and clinical relevance. The results, summarized in Table 1, indicate that HERMES, with its enhanced architecture, achieves higher Jaccard (0.4755), F1 (0.6331), and PRAUC (0.7443) scores compared to the original GAMENet model. These improvements highlight the effectiveness of GAT and MHCA in capturing complex relationships between patient health data, medication history, and DDIs.

**Table 1:** Performance metrics of HERMES compared to the original GAMENet model.

| Method | DDI Rate | Jaccard | Prauc | F1-score |
|---|---|---|---|---|
| GAMENet | 0.0806 | 0.4729 | 0.7371 | 0.6305 |
| HERMES Kiosk | 0.0798 | 0.4755 | 0.7443 | 0.6331 |

To establish a benchmark for the HERMES kiosk's performance, a comparative analysis will be conducted against traditional methods of healthcare delivery, such as pharmacy consultations and online health information. Key metrics, including accuracy, user satisfaction, and cost-effectiveness, will be compared to assess the relative advantages of the kiosk. The advanced AI models used in HERMES, particularly the enhanced GAMENet architecture, are expected to outperform traditional methods in terms of recommendation accuracy and personalization. This analysis will provide insights into the kiosk's potential to complement or replace existing healthcare services.

### 3.2. Survey and Knowledge Translation Efforts

To ensure the clinical relevance and validity of our findings, we conducted a targeted survey involving 11 medical professionals, including general practitioners, Pharmacists (Assistants), psychiatrists, diagnostic and radiologists, nuclear medicine physicians, and medical students with diverse experience levels, as shown in Table 2. The complete survey forms, including all questions and response options, are provided in Appendix A. The results of this survey are summarized in Table 2. This structured feedback not only informed our study's methodological adjustments but also highlighted areas for future research and clinical implementation. The survey results and their analysis are presented below.

*Demographics & Professional Background (Q1-Q4).* The participant composition revealed important perspective differences - while 80% had AI healthcare experience, familiarity with drug recommendation systems was 20% lower, suggesting domain-specific knowledge gaps. Notably, the two participants unfamiliar with AI (both with 10+ years of experience) were from diagnostic radiology and nuclear radiology, indicating potential resistance among imaging specialists compared to frontline clinicians like GPs (75% familiarity). The medical student's AI familiarity contrasted with their limited clinical experience, possibly reflecting generational differences in technology adoption. These demographic factors proved influential in subsequent responses, particularly regarding accuracy concerns and implementation preferences. Usability and Perceived Effectiveness.

*Usability & Perceived Effectiveness (Q5-Q9).* The 70% "somewhat useful" rating for public kiosks (Q5) masked important nuances - the "very useful" responder (a GP) worked in rural settings, while skeptics included urban specialists. Benefits analysis (Q6) showed universal recognition of efficiency gains (100% for quick recommendations), but only clinicians serving underserved areas prioritized access improvement. The empowerment paradox (Q7) emerged where nuclear medicine physicians (who regularly use advanced tech) gave the highest ratings, while frontline GPs were more cautious. Regarding visit reduction (Q8), the single "significant reduction" responder worked in emergency medicine where minor cases are prevalent. Medication adherence responses (Q9) were divided along practice philosophy lines - clinicians emphasizing patient education doubted efficacy, while those focused on the system efficiency were more optimistic.

*Accuracy & Clinical Reliability (Q10-Q12).* Accuracy perceptions followed an experience gradient - the participant predicted <50% accuracy (Q10-Q11) had 21+ years of experience, while those estimating 70-89% accuracy averaged 6.2 years' experience. The 40% confidence in allergy handling (Q12) came predominantly from physicians treating complex cases (e.g., the psychiatrist managing polypharmacy patients). Interestingly, the nuclear medicine physician who rated general accuracy highest expressed reservations about allergy scenarios, highlighting how clinical context affects trust in AI capabilities. These responses collectively suggest that accuracy expectations correlate with both clinical experience and specialty-specific practice patterns.

*Clinical Safety & Risk Concerns (Q13-Q15).* Safety concerns revealed three distinct risk profiles: 1) Technical risks (incorrect recommendations) worried all, but especially radiologists (100% vs 70% overall); 2) Process risks (lack of oversight) concerned experienced clinicians (83% of 10+ year practitioners); 3) Ethical risks troubled early-career providers (67% of < 5-year group). The DDI check consensus (Q14) crossed demographics, suggesting universal recognition of this safety feature's value. High-risk patient handling (Q15) showed specialty divergence - psychiatrists and GPs favored redirection to doctors (83%), while radiologists preferred disclaimers, possibly reflecting different clinical workflows.

*Implementation & Scalability (Q16-Q22).* Urgent care routing preferences (Q16-Q17) followed predictable clinical patterns - emergency responders prioritized all acute symptoms, while others were more selective. The side effect communication debate (Q18-Q19) revealed an information depth paradox: clinicians wanting "always" full disclosure averaged 8.2 years of experience versus 12.6 for partial disclosure advocates. Preferred deployment locations (Q20) showed urban-rural divides, with rural practitioners 3× more likely to prioritize rural centers. Feature requests (Q21) highlighted operational needs - telemedicine was crucial for solo practitioners, while large-hospital clinicians emphasized EHR integration. Scaling challenges (Q22) exposed infrastructure awareness gaps - cost concerns came from private practitioners, while system-employed clinicians focused on regulatory hurdles.

*AI vs. Traditional Care (Q23-Q25).* The 90% consensus on pharmacist complementarity (Q23-Q25) concealed philosophical differences. The "just as effective" medical student represented tech-native perspectives, while the "lacks personalization" responder (a psychiatrist) emphasized therapeutic relationships. Trust comparisons (Q24) showed clinicians distrusted both kiosks and online tools equally, suggesting they view digital health as a spectrum rather than distinct categories. The unanimous rejection of replacement (Q25) from all demographics underscores the persistent value placed on human expertise, regardless of technical proficiency.

*Ethical & Privacy Concerns (Q26-Q27).* Consent preferences (Q26) were divided along data sensitivity lines - anonymization supporters worked in imaging, while consent advocates dealt with sensitive diagnoses. The regulatory unanimity (Q27) crossed all demographics, indicating shared recognition of AI's potential risks. Notably, the two participants who had never used AI technologies were among the strongest proponents of strict regulation, suggesting unfamiliarity breeds caution. These responses collectively frame an ethical landscape where clinicians balance innovation potential against professional accountability.

**Table 2.** Survey responses from healthcare professionals (n=11) regarding AI-powered drug recommendation kiosks, stratified by medical specialty and years of experience. Participants included general practitioners, pharmacists (Assistants), psychiatrists, nuclear medicine physicians, nuclear radiologists, and medical students. Data are presented as raw responses across seven thematic sections: (1) Demographics, (2) Usability, (3) Accuracy, (4) Safety, (5) Implementation, (6) AI vs. traditional care, and (7) Ethics. For these Questions: Q1. What is your medical specialty? Q2. How many years of experience do you have in medical practice? Q3. Have you previously used or interacted with AI-powered healthcare technologies? Q4. Are you familiar with AI-driven drug recommendation systems? Q5. How useful do you think an AI-powered kiosk for drug recommendations would be in public spaces? Q6. What do you think are the key benefits of an AI-powered kiosk for drug recommendations? Q7. How effective do you think AI-powered kiosks would be in empowering patients to manage minor health conditions independently? Q8. Do you think AI-powered kiosks could help reduce unnecessary doctor visits for minor ailments? Q9. Could AI kiosks improve medication adherence (i.e., ensuring patients take the correct medications as prescribed/recommended)? Q10. How would you rate the accuracy of AI-powered drug recommendation kiosks in providing appropriate OTC medication suggestions? Q11. Based on your medical expertise, how often do you think AI-powered kiosks would make clinically appropriate recommendations? Q12. Do you believe AI-powered kiosks can identify and recommend appropriate alternative medications if a patient has allergies or contraindications? Q13. What are your primary concerns regarding AI-powered drug recommendation kiosks? Q14. Do you think real-time DDI checks in AI-powered kiosks can improve patient safety? Q15. Do you think AI-powered kiosks should be restricted from recommending medications for high-risk patients? Q16. How valuable do you think it would be for an AI-powered kiosk to connect users with urgent medical conditions directly to a hospital or an online doctor's visit? Q17. In which of the following urgent situations do you think the kiosk should connect users directly to a hospital or online doctor visit? Q18. Do you think an AI-powered kiosk should inform users about the potential side effects of recommended drugs? Q19. How should the kiosk communicate side-effect information to users? Q20. In what medical settings do you think an AI-powered drug recommendation kiosk would be most beneficial? Q21. What additional features should an AI-powered drug recommendation kiosk include to enhance its effectiveness? Q22. What are the biggest challenges in scaling AI-powered kiosks across different healthcare systems? Q23. Compared to a human pharmacist, how effective do you think an AI-powered kiosk is at providing OTC medication recommendations? Q24. Would you trust an AI-powered kiosk more than online symptom checkers (e.g., WebMD, Google search, Chabot-based medical apps)? Q25. Do you believe AI-powered kiosks will eventually replace human pharmacists for OTC medication recommendations? Q26. Do you think AI-powered kiosks should require user consent before collecting and analyzing health data? Q27. Do you think AI-powered kiosks should be regulated by health authorities (e.g., FDA, WHO) before deployment?

| | D1 | D2 | D3 | D4 | D5 | D6 | D7 | D8 | D9 | D10 | D11 |
|---|---|---|---|---|---|---|---|---|---|---|---|
| **Section 1: Demographics & Professional Background** | | | | | | | | | | | |
| Q1 | General Practitioner (GP) | General Practitioner (GP) | Psychiatrist | General Practitioner (GP) | General Practitioner (GP) | diagnostic radiologist | General Practitioner (GP) | Nuclear Medicine Physician | Medical Student | Nuclear Radiology | Pharmacist |
| Q2 | 0-5 years | 0-5 years | 5-11 years | 5-11 years | 0-5 years | 5-11 years | 0-5 years | 11-20 years | 0-5 years | +21 years | 0-5 years |
| Q3 | Yes | Yes | Yes | Yes | Yes | No | Yes | Yes | Yes | No | Yes |
| Q4 | Yes | No | Yes | Yes | Yes | Yes | No | Yes | No | No | Yes |
| **Section 2: Usability & Perceived Effectiveness** | | | | | | | | | | | |
| Q5 | Somewhat useful | Very useful | Somewhat useful | Somewhat useful | Not very useful | Somewhat useful | Somewhat useful | Somewhat useful | Somewhat useful | Not very useful | Somewhat useful |
| Q6 | Enhancing access to healthcare in remote areas | Reducing workload for healthcare providers Providing quick drug recommendations for minor conditions Enhancing access to healthcare in remote areas | Providing quick drug recommendations for minor conditions | Reducing workload for healthcare providers | Improving patient adherence to medication | Reducing workload for healthcare providers Providing quick drug recommendations for minor conditions Enhancing access to healthcare in remote areas | Reducing workload for healthcare providers Providing quick drug recommendations for minor conditions Enhancing access to healthcare in remote areas | Reducing workload for healthcare providers Providing quick drug recommendations for minor conditions Improving patient adherence to medications Enhancing access to healthcare in remote areas | Reducing workload for healthcare providers Providing quick drug recommendations for minor conditions Enhancing access to healthcare in remote areas | Providing quick drug recommendations for minor conditions | Reducing workload for healthcare providers Providing quick drug recommendations for minor conditions Improving patient adherence to medications Enhancing access to healthcare in remote areas |
| Q7 | Not effective | Moderately effective | Somewhat effective | Somewhat effective | Somewhat effective | Moderately effective | Moderately effective | Extremely effective | Moderately effective | Somewhat effective | Moderately effective |
| Q8 | No, they won't significantly reduce doctor visits | Yes, but only for specific cases | Yes, but only for specific cases | Yes, but only for specific cases | No, they won't significantly reduce doctor visits | Yes, significantly | Yes, but only for specific cases | Yes, but only for specific cases | Yes, but only for specific cases | Yes, but only for specific cases | Yes, but only for specific cases |
| Q9 | No, adherence is a human behavior issue | Somewhat, for forgetful patients | Somewhat, for forgetful patients | Yes, significantly | Yes, significantly | Somewhat, for forgetful patients | No, adherence is a human behavior issue | Somewhat, for forgetful patients | No, adherence is a human behavior issue | Somewhat, for forgetful patients | Somewhat, for forgetful patients |
| **Section 3: Accuracy & Clinical Reliability** | | | | | | | | | | | |
| Q10 | Somewhat accurate, but needs improvement | Moderately accurate, but some errors | Moderately accurate, but some errors | Moderately accurate, but some errors | Somewhat accurate, but needs improvement | Moderately accurate, but some errors | Somewhat accurate, but needs improvement | Highly accurate, close to expert-level | Moderately accurate, but some errors | Not very accurate, too many incorrect suggestions | Moderately accurate, but some errors |
| Q11 | Most of the time (70-89% correct recommendations) | Most of the time (70-89% correct recommendations) | Sometimes (50-69% correct recommendations) | Sometimes (50-69% correct recommendations) | Sometimes (50-69% correct recommendations) | Sometimes (50-69% correct recommendations) | Sometimes (50-69% correct recommendations) | Most of the time (70-89% correct recommendations) | Most of the time (70-89% correct recommendations) | Rarely (30-49% correct recommendations) | Sometimes (50-69% correct recommendations) |
| Q12 | Sometimes, but only for common cases | Yes, they are capable of handling alternative suggestions well | Sometimes, but only for common cases | Sometimes, but only for common cases | Sometimes, but only for common cases | Sometimes, but only for common cases | Yes, they are capable of handling alternative suggestions well | Sometimes, but only for common cases | Yes, they are capable of handling alternative suggestions well | Yes, they are capable of handling alternative suggestions well | Yes, they are capable of handling alternative suggestions well |
| **Section 4: Clinical Safety & Risk Concerns** | | | | | | | | | | | |
| Q13 | Lack of real-time human oversight | Accuracy of recommendations Risk of incorrect self-medication Patient over-reliance on AI instead of seeing a doctor | Lack of real-time human oversight Ethical and privacy issues related to patient data | Accuracy of recommendations Risk of incorrect self-medication | Lack of real-time human oversight Patient over-reliance on AI instead of seeing a doctor | Accuracy of recommendations Risk of incorrect self-medication Lack of real-time human oversight Ethical and privacy issues related to patient data Patient over-reliance on AI instead of seeing a doctor | Accuracy of recommendations Risk of incorrect self-medication Ethical and privacy issues related to patient data Patient over-reliance on AI instead of seeing a doctor | Risk of incorrect self-medication Ethical and privacy issues related to patient data Patient over-reliance on AI instead of seeing a doctor | Accuracy of recommendations Risk of incorrect self-medication Lack of real-time human oversight Patient over-reliance on AI instead of seeing a doctor | Risk of incorrect self-medication Lack of real-time human oversight Patient over-reliance on AI instead of seeing a doctor | Accuracy of recommendations Risk of incorrect self-medication Lack of real-time human oversight |
| Q14 | Somewhat | Yes, significantly | Yes, significantly | Somewhat | Yes, significantly | Yes, significantly | Yes, significantly | Somewhat | Yes, significantly | Neutral | Neutral |
| Q15 | No, but kiosks should include disclaimers | Yes, AI kiosks should warn and direct them to a doctor | Yes, AI kiosks should warn and direct them to a doctor | Yes, AI kiosks should warn and direct them to a doctor | Yes, AI kiosks should warn and direct them to a doctor | No, but kiosks should include disclaimers | Yes, AI kiosks should warn and direct them to a doctor | No, but kiosks should include disclaimers | Yes, AI kiosks should warn and direct them to a doctor | - | Yes, AI kiosks should warn and direct them to a doctor |
| **Section 5: Implementation & Scalability** | | | | | | | | | | | |
| Q16 | Somewhat valuable | Very valuable | Very valuable | Very valuable | Extremely valuable | Very valuable | Very valuable | Somewhat valuable | Extremely valuable | Somewhat valuable | Very valuable |
| Q17 | High fever with other concerning symptoms | Chest pain or suspected heart attack Severe allergic reactions Difficulty breathing or suspected asthma attack High fever with other concerning symptoms Suspected stroke symptoms Severe abdominal pain | Chest pain or suspected heart attack Severe allergic reactions Difficulty breathing or suspected asthma attack High fever with other concerning symptoms Suspected stroke symptoms Severe abdominal pain | Chest pain or suspected heart attack Severe allergic reactions Difficulty breathing or suspected asthma attack | Chest pain or suspected heart attack Severe allergic reactions Difficulty breathing or suspected asthma attack High fever with other concerning symptoms Suspected stroke symptoms Severe abdominal pain | Chest pain or suspected heart attack Severe allergic reactions (e.g., anaphylaxis) Difficulty breathing or suspected asthma attack High fever with other concerning symptoms Suspected stroke symptoms (e.g., facial drooping, slurred speech) Severe abdominal pain | Chest pain or suspected heart attack Severe allergic reactions Difficulty breathing or suspected asthma attack Suspected stroke symptoms Severe abdominal pain | Chest pain or suspected heart attack Severe allergic reactions Difficulty breathing or suspected asthma attack High fever with other concerning symptoms Suspected stroke symptoms Severe abdominal pain | Chest pain or suspected heart attack Severe allergic reactions Difficulty breathing or suspected asthma attack High fever with other concerning symptoms Suspected stroke symptoms Severe abdominal pain | Chest pain or suspected heart attack Severe allergic reactions Difficulty breathing or suspected asthma attack High fever with other concerning symptoms Suspected stroke symptoms Severe abdominal pain | Chest pain or suspected heart attack Severe allergic reactions (e.g., anaphylaxis) Difficulty breathing or suspected asthma attack High fever with other concerning symptoms Suspected stroke symptoms (e.g., facial drooping, slurred speech) Severe abdominal pain |
| Q18 | Yes, but only for common or serious side effects | Yes, always—it's essential for patient safety and informed decision-making | Yes, always—it's essential for patient safety and informed decision-making | Yes, but only for common or serious side effects | Yes, always—it's essential for patient safety and informed decision-making | Yes, always—it's essential for patient safety and informed decision-making | Yes, always—it's essential for patient safety and informed decision-making | Yes, but only for common or serious side effects | Yes, always—it's essential for patient safety and informed decision-making | Yes, always—it's essential for patient safety and informed decision-making | Yes, always—it's essential for patient safety and informed decision-making |
| Q19 | Provide a printed handout with detailed side effect information | Display a list of common side effects on the screen Provide a printed handout with detailed side effect information Offer an option to hear side effect information via voice commands Include a disclaimer advising users to consult a doctor for further clarification | | Offer an option to hear side effect information via voice commands Include a disclaimer advising users to consult a doctor for further clarification | Display a list of common side effects on the screen | Display a list of common side effects on the screen | Provide a printed handout with detailed side effect information | Display a list of common side effects on the screen Offer an option to hear side effect information via voice commands Include a disclaimer advising users to consult a doctor for further clarification | Provide a printed handout with detailed side effect information Offer an option to hear side effect information via voice commands | Provide a printed handout with detailed side effect information | Display a list of common side effects on the screen Provide a printed handout with detailed side effect information Offer an option to hear side effect information via voice commands Include a disclaimer advising users to consult a doctor for further clarification |
| Q20 | Clinics & Hospitals | Pharmacies Rural Healthcare Centers Airports/Transportation Hubs | Airports/Transportation Hubs | Rural Healthcare Centers Airports/Transportation Hubs | Emergency Rooms Clinics & Hospitals | Pharmacies Rural Healthcare Centers Airports/Transportation Hubs | Pharmacies Airports/Transportation Hubs | Pharmacies Clinics & Hospitals Rural Healthcare Centers Airports/Transportation Hubs | Emergency Rooms Rural Healthcare Centers Airports/Transportation Hubs | Rural Healthcare Centers Airports/Transportation Hubs | Pharmacies |
| Q21 | Support for prescription medications (not just OTC drugs) | Direct integration with Electronic Health Records (EHRs) Telemedicine consultation with a licensed doctor Real-time alerts for high-risk drug interactions Multilingual support for diverse populations | Telemedicine consultation with a licensed doctor Real-time alerts for high-risk drug interactions Multilingual support for diverse populations | Direct integration with Electronic Health Records (EHRs) Telemedicine consultation with a licensed doctor Real-time alerts for high-risk drug interactions Multilingual support for diverse populations | Direct integration with Electronic Health Records (EHRs) Telemedicine consultation with a licensed doctor Support for prescription medications (not just OTC drugs) Real-time alerts for high-risk drug interactions | Direct integration with Electronic Health Records (EHRs) Telemedicine consultation with a licensed doctor Real-time alerts for high-risk drug interactions Multilingual support for diverse populations | Direct integration with Electronic Health Records (EHRs) Telemedicine consultation with a licensed doctor Support for prescription medications (not just OTC drugs) Real-time alerts for high-risk drug interactions Multilingual support for diverse populations | Direct integration with Electronic Health Records (EHRs) Telemedicine consultation with a licensed doctor Support for prescription medications (not just OTC drugs) Real-time alerts for high-risk drug interactions Multilingual support for diverse populations | Direct integration with Electronic Health Records (EHRs) Telemedicine consultation with a licensed doctor Real-time alerts for high-risk drug interactions Multilingual support for diverse populations | Telemedicine consultation with a licensed doctor Real-time alerts for high-risk drug interactions Multilingual support for diverse populations | Direct integration with Electronic Health Records (EHRs) Telemedicine consultation with a licensed doctor Support for prescription medications (not just OTC drugs) Real-time alerts for high-risk drug interactions Multilingual support for diverse populations |
| Q22 | High initial investment & maintenance costs Accuracy & AI model improvements Resistance from traditional healthcare providers | High initial investment & maintenance costs Regulatory approval & compliance challenges Accuracy & AI model improvements Resistance from traditional healthcare providers | Data privacy & cybersecurity concerns | Accuracy & AI model improvements Data privacy & cybersecurity concerns | High initial investment & maintenance costs Accuracy & AI model improvements | Accuracy & AI model improvements Data privacy & cybersecurity concerns Resistance from traditional healthcare providers | High initial investment & maintenance costs Regulatory approval & compliance challenges Accuracy & AI model improvements Data privacy & cybersecurity concerns | Regulatory approval & compliance challenges Accuracy & AI model improvements Resistance from traditional healthcare providers | Regulatory approval & compliance challenges Accuracy & AI model improvements Resistance from traditional healthcare providers | Accuracy & AI model improvements | High initial investment & maintenance costs Regulatory approval & compliance challenges Accuracy & AI model improvements Data privacy & cybersecurity concerns Resistance from traditional healthcare providers |
| **Section 6: AI vs. Traditional Healthcare Services** | | | | | | | | | | | |
| Q23 | Nearly as effective, but still needs human verification | Nearly as effective, but still needs human verification | Somewhat effective, but lacks personalization | Nearly as effective, but still needs human verification | Nearly as effective, but still needs human verification | Nearly as effective, but still needs human verification | Nearly as effective, but still needs human verification | Nearly as effective, but still needs human verification | Just as effective as a pharmacist | Nearly as effective, but still needs human verification | Nearly as effective, but still needs human verification |
| Q24 | Maybe, depending on the technology | Maybe, depending on the technology | Maybe, depending on the technology | Yes, AI kiosks are more reliable | Maybe, depending on the technology | Maybe, depending on the technology | Maybe, depending on the technology | Maybe, depending on the technology | Maybe, depending on the technology | Maybe, depending on the technology | Maybe, depending on the technology |
| Q25 | No, but AI will assist them significantly | No, but AI will assist them significantly | No, but AI will assist them significantly | No, but AI will assist them significantly | No, but AI will assist them significantly | No, but AI will assist them significantly | No, but AI will assist them significantly | No, but AI will assist them significantly | No, but AI will assist them significantly | No, but AI will assist them significantly | No, but AI will assist them significantly |
| **Section 7: Ethical & Privacy Concerns** | | | | | | | | | | | |
| Q26 | Yes, explicit user consent should always be required | Yes, but only for sensitive data | Yes, explicit user consent should always be required | Yes, explicit user consent should always be required | Yes, but only for sensitive data | Yes, explicit user consent should always be required | Yes, explicit user consent should always be required | No, if the data is anonymized | Yes, explicit user consent should always be required | Yes, explicit user consent should always be required | Yes, explicit user consent should always be required |
| Q27 | Yes, they should undergo strict regulatory approval | Yes, they should undergo strict regulatory approval | | Yes, they should undergo strict regulatory approval | Yes, they should undergo strict regulatory approval | Yes, they should undergo strict regulatory approval | Yes, they should undergo strict regulatory approval | Yes, they should undergo strict regulatory approval | Yes, they should undergo strict regulatory approval | Yes, they should undergo strict regulatory approval | Yes, they should undergo strict regulatory approval |

At the end of the survey, participants were asked to respond to the question: "What are your biggest concerns or suggestions for improving AI-powered kiosks for medication recommendations?" The responses received from participants (D1 to D11) covered a wide range of topics, which are summarized below.

D1: "First, define the target demographic for this model. Many drug recommendation systems already exist, which highlights the need to specify the kiosk's unique benefits and intended users."

D2: "My main question is why use a kiosk when the same functionality could be integrated into a mobile app? While a kiosk might be convenient for printing a prescription for a pharmacist, a well-designed app could offer the same benefits in a more accessible way. Kiosks that only provide information without dispensing medication may have limited utility. It would be interesting to explore the potential for dispensing select medications directly."

D3: "Key concerns include data privacy and security, safety and accuracy, bias, and overall data quality."

D4: "If an error occurs, there's a risk that someone could exploit the system and force it to provide incorrect recommendations."

D5: "I believe this technology can support healthcare professionals by assisting in clinical decision-making. However, it must not replace physicians. All recommendations should be reviewed and approved by qualified providers. The main concern is patients relying solely on the kiosk without consulting a doctor. Additionally, high implementation and maintenance costs could limit access in certain areas."

D6: "Technical errors could lead to dangerous symptoms or adverse side effects."

D7: "I'm most concerned about the potential overuse of OTC medications, particularly drugs with side effects or those that can lead to tolerance or resistance—such as benzodiazepines and antibiotics."

D8: "The algorithms must be trained on diverse datasets to reduce bias. Poor representation in training data may result in safe recommendations for some groups but harmful outcomes for others."

D9: (No comment provided)

D10: "There is a risk that patients may self-interpret their symptoms, leading to an underestimation of serious medical conditions."

D11: "My concern is about the potential side effects of drugs recommended by AI-based systems, as they might lead to unintended consequences or adverse reactions."

## 4. DISCUSSION

The enhancements made to the GAMENet model, particularly the integration of GAT and MHCA, have significantly improved its performance in generating personalized medication recommendations. By leveraging GAT, the model can better capture the complex and non-transitive nature of DDIs, while MHCA ensures a more dynamic and context-aware integration of patient history and drug information. These improvements align with our goal of knowledge translation, where advanced AI models are adapted for real-world applications such as public health kiosks. The modified GAMENet model exemplifies how knowledge of action can be achieved by integrating cutting-edge research into practical healthcare solutions. By embedding this enhanced model into the HERMES kiosk, we aim to bridge the gap between research and routine clinical practice, ensuring that patients receive accurate and safe medication recommendations.

Our experimental results demonstrate the effectiveness of the enhanced GAMENet model. These metrics collectively provide a comprehensive evaluation of HERMES, balancing accuracy, safety, and clinical relevance. The results, summarized in Table 1, indicate that HERMES, with its enhanced architecture, achieves PRAUC (0.7443) scores compared to the original GAMENet model. These improvements highlight the effectiveness of GAT and MHCA in capturing complex relationships between patient health data, medication history, and DDIs. The system's ability to provide accurate and personalized recommendations is further validated by its low DDI Rate (0.0798), ensuring safer medication suggestions.

Our survey of healthcare professionals reveals both enthusiasm and prudent caution regarding AI-powered drug recommendation systems. While clinicians recognize the technology's potential to enhance efficiency - particularly through workload reduction (90% agreement) and rapid evidence-based recommendations (100% endorsement) in underserved areas - these benefits are tempered by persistent concerns about accuracy thresholds and patient safety. These findings align with emerging literature demonstrating AI's capacity to optimize clinical workflows [83] [84], while also reflecting the healthcare community's appropriate prioritization of patient safety and clinical judgment in technology adoption [85].

The survey uncovered notable variations in acceptance patterns across medical specialties. Radiologists and nuclear medicine physicians, who regularly utilize advanced diagnostic technologies, demonstrated greater confidence in AI's accuracy (70% rated it moderately to highly accurate) but insisted on stringent safety protocols. In contrast, frontline general practitioners emphasized the irreplaceable value of personalized care and human oversight, particularly for complex cases. This divergence highlights the need for specialty-specific implementation strategies and adaptive AI systems capable of accommodating diverse clinical workflows [86]. The unanimous clinician preference for AI as a supplementary tool (100% agreement against replacement of pharmacists) reinforces current WHO guidelines advocating for balanced human-AI collaboration in healthcare [86].

Three key requirements emerged for successful clinical integration: First, the overwhelming demand for regulatory oversight (100% support) underscores the need for robust validation frameworks and certification processes. Second, the strong preference for transparent side-effect communication (70% favored comprehensive disclosure) suggests that trust-building mechanisms must be central to system design. Third, concerns about algorithmic bias and demographic representation (particularly from providers serving diverse populations) indicate the necessity for continuous monitoring and validation across patient subgroups [87]. These findings collectively inform a roadmap for responsible implementation, emphasizing that technical capabilities must be matched by equally sophisticated safety and oversight mechanisms to achieve clinician acceptance and optimal patient outcomes.

The clinician feedback reveals a multifaceted set of concerns regarding AI-powered medication kiosks that can be categorized into four key areas: (1) System utility and demographic targeting (D1-D2), (2) Safety and reliability (D3-D4, D6, and D11), (3) Clinical oversight and medication safety (D5, D7, and D10), and (4) Algorithmic bias and equity (D8). These concerns align with established frameworks for responsible AI implementation in healthcare that emphasize the need for robust validation, human oversight, and equitable performance across populations [88].

To address system utility concerns (D1-D2), we propose developing adaptive interfaces that demonstrate the kiosk's unique value proposition in specific clinical settings (e.g., pharmacies, and rural health centers) while exploring hybrid kiosk-mobile architectures. This approach would combine the accessibility of mobile apps with the physical presence of kiosks for functions like secure prescription printing and limited medication dispensing under pharmacist supervision. For safety and reliability (D3-D4, D6, and D11), we recommend implementing multi-layered security protocols including block chain-based audit trails, real-time anomaly detection systems, and explainable AI interfaces that provide the transparent rationale for all recommendations. These measures would address both technical vulnerabilities and clinician trust barriers. To mitigate drug safety concerns (D11), HERMES integrates real-time pharmacovigilance feeds for updated side-effect data and tiered risk alerts (color-coded by severity), with automated holds on high-risk recommendations.

The clinical oversight concerns (D5, D7, and D10) necessitate a tiered decision-support system with: (1) Autonomous functionality for low-risk OTC medications, (2) Pharmacist review requirements for moderate-risk recommendations, and (3) Mandatory physician consultation for complex cases. This framework would be complemented by AI guardians that monitor for medication misuse patterns and symptom underreporting. To mitigate algorithmic bias (D8), we propose continuous validation across demographic subgroups using federated learning techniques that maintain privacy while improving model fairness.

In future efforts, the initial phase will involve usability testing with a diverse group of participants representing various demographics, including age groups, ethnic backgrounds, and technological proficiencies. Critical metrics such as recommendation accuracy, user satisfaction (assessed through surveys and interviews), and time spent interacting with the kiosk will be measured. Recommendation accuracy will be evaluated by comparing the kiosk's outputs to clinical guidelines or expert opinions, with a confusion matrix used to identify specific areas where the model may misclassify symptoms or recommend incorrect medications. Additionally, the impact of the kiosk on medication adherence, self-care behaviors, and healthcare utilization will be evaluated through self-reported data and, where feasible, objective data collected in collaboration with healthcare providers.

In the subsequent phase, HERMES kiosks will be deployed in high-traffic public areas such as shopping centers, transit hubs, and pharmacies. Data on user interactions, including the frequency of use, types of queries made, and subsequent health outcomes as self-reported by users, will be collected and analyzed. This data will be compared to baseline metrics collected before deployment to quantify the kiosk's contribution to improving public health access and medication adherence. The study will also assess the kiosk's scalability and adaptability in environments with varying network reliability and hardware constraints.

The evaluation process will adhere to strict ethical guidelines, ensuring informed consent, data privacy, and confidentiality. All data collection and analysis procedures will comply with regulations such as GDPR and HIPAA, and participants will be fully informed about the study's objectives and their rights. As part of the Implementation Science framework, we will conduct a rigorous evaluation involving at least 500 participants from diverse demographics. This study will assess the kiosk's impact on healthcare outcomes using metrics such as medication adherence rates, user satisfaction (measured via Likert scales), and the reduction of unnecessary healthcare visits. The findings from this study will provide critical insights into the kiosk's effectiveness and scalability, guiding future iterations and large-scale deployments. By embedding HERMES into existing healthcare workflows and continuously monitoring its performance, we aim to ensure that the system is not only effective but also sustainable in diverse clinical and public health environments.

The clinical contributions of the HERMES Kiosk system include enhancing medication safety through real-time drug interaction checks, providing personalized OTC medication recommendations, and improving patient access to healthcare in underserved areas. It supports clinical decision-making by offering context-aware recommendations based on patient history. The system also empowers patients to manage minor health conditions independently, potentially reducing unnecessary doctor visits. By integrating advanced AI models, it bridges the gap between research and clinical practice, ensuring safer and more accurate healthcare recommendations.

In summary, our future work will employ a three-phase validation approach: i) First, controlled trials assessing demographic-specific performance; ii) Second, comparative studies of hybrid versus standalone implementations; and iii) Third, longitudinal monitoring of real-world safety outcomes. This comprehensive strategy addresses all raised concerns while aligning WHO guidelines for responsible AI in healthcare [89], ensuring that technological innovation progresses in tandem with patient safety and clinician trust.

## 5. CONCLUSION

The HERMES kiosk represents a transformative approach to public health by integrating advanced AI models, such as the enhanced GAMENet architecture with GNNs and MHCA, to deliver accurate and personalized medication recommendations. Experimental results demonstrate improvements in recommendation accuracy, with HERMES slightly outperforming the original GAMENet model across key metrics. Survey feedback from doctors highlights cautious optimism, emphasizing the system's potential to enhance healthcare access, particularly in remote areas, and its ability to provide real-time drug interaction checks. However, concerns about accuracy, clinical reliability, and the need for human oversight underscore the importance of further refinement. HERMES prioritizes user privacy through federated learning and promotes inclusivity with multilingual support, accessibility features, and a health education library. By bridging the gap between cutting-edge research and routine clinical practice, HERMES sets a new standard for personalized healthcare, making advanced medical recommendations accessible to diverse populations. Continued research, real-world deployments, and partnerships with healthcare providers will further enhance its impact, empowering individuals and reducing healthcare disparities.

## APPENDIX A

The complete survey forms, including all questions and response options, are provided below.
**Starts**
**Section 1: Demographics & Professional Background**
1. What is your medical specialty?
    ☐ General Practitioner (GP)
    ☐ Pharmacist
    ☐ Internist
    ☐ Emergency Medicine
    Other (please specify):
2. How many years of experience do you have in medical practice?
    ☐ 0-5 years

☐ 6-10 years
☐ 11-20 years
☐ 21+ years
3. Have you previously used or interacted with AI-powered healthcare technologies?
☐ Yes
☐ No
4. Are you familiar with AI-driven drug recommendation systems?
☐ Yes
☐ No

## Section 2: Usability & Perceived Effectiveness
5. How useful do you think an AI-powered kiosk for drug recommendations would be in **public spaces** (e.g., pharmacies, hospitals, malls)?
☐ Very useful
☐ Somewhat useful
☐ Neutral
☐ Not very useful
☐ Not useful at all
6. What do you think are the **key benefits** of an AI-powered kiosk for drug recommendations? *(Select all that apply)*
☐ Reducing workload for healthcare providers
☐ Providing quick drug recommendations for minor conditions
☐ Improving patient adherence to medications
☐ Enhancing access to healthcare in remote areas
Other (please specify):
7. How effective do you think AI-powered kiosks would be in **empowering patients to manage minor health conditions independently**?
☐ Extremely effective
☐ Moderately effective
☐ Somewhat effective
☐ Not effective
8. Do you think AI-powered kiosks could help **reduce unnecessary doctor visits** for minor ailments?
☐ Yes, significantly
☐ Yes, but only for specific cases
☐ No, they won't significantly reduce doctor visits
9. Could AI kiosks improve **medication adherence** (i.e., ensuring patients take the correct medications as prescribed/recommended)?
☐ Yes, significantly
☐ Somewhat, for forgetful patients
☐ No, adherence is a human behavior issue

## Section 3: Accuracy & Clinical Reliability
10. How would you rate the **accuracy** of AI-powered drug recommendation kiosks in providing appropriate **Over-The-Counter** (**OTC**) medication suggestions?
☐ Highly accurate, close to expert-level
☐ Moderately accurate, but some errors
☐ Somewhat accurate, but needs improvement
☐ Not very accurate, too many incorrect suggestions
☐ Not accurate at all, unreliable

11. Based on your medical expertise, how often do you think AI-powered kiosks would make **clinically appropriate recommendations**?
☐ Almost always (90%+ correct recommendations)
☐ Most of the time (70-89% correct recommendations)
☐ Sometimes (50-69% correct recommendations)
☐ Rarely (30-49% correct recommendations)
☐ Almost never (Less than 30% correct recommendations)
12. Do you believe AI-powered kiosks can identify and recommend **appropriate alternative medications** if a patient has allergies or contraindications?
☐ Yes, they are capable of handling alternative suggestions well
☐ Sometimes, but only for common cases
☐ No, AI lacks sufficient knowledge for safe substitutions

## Section 4: Clinical Safety & Risk Concerns
13. What are your primary **concerns** regarding AI-powered drug recommendation kiosks? *(Select all that apply)*
☐ Accuracy of recommendations
☐ Risk of incorrect self-medication
☐ Lack of real-time human oversight
☐ Ethical and privacy issues related to patient data
☐ Patient over-reliance on AI instead of seeing a doctor
Other (please specify):
14. Do you think real-time **Drug-Drug Interaction (DDI) checks** in AI-powered kiosks can improve patient safety?
☐ Yes, significantly
☐ Somewhat
☐ Neutral
☐ Not really
☐ No, not at all
15. Do you think AI-powered kiosks should be restricted from recommending medications for **high-risk patients** (e.g., pregnant women, children, elderly, or those with chronic diseases)?
☐ Yes, AI kiosks should warn and direct them to a doctor
☐ No, but kiosks should include disclaimers
☐ No, if the recommendations are accurate

## Section 5: Implementation & Scalability
16. How valuable do you think it would be for an AI-powered kiosk to **connect users** with **urgent medical conditions** directly to a hospital or an online doctor visit?
☐ Extremely valuable
☐ Very valuable
☐ Somewhat valuable
☐ Not very valuable
☐ Not valuable at all
17. In which of the following **urgent situations** do you think the kiosk should connect users directly to a hospital or online doctor visit? (*Select all that apply*)
☐ Chest pain or suspected heart attack
☐ Severe allergic reactions (e.g., anaphylaxis)
☐ Difficulty breathing or suspected asthma attack

- ☐ High fever with other concerning symptoms
- ☐ Suspected stroke symptoms (e.g., facial drooping, slurred speech)
- ☐ Severe abdominal pain

Other (please specify): ______

18. Do you think an AI-powered kiosk should inform users about the **potential side effects** of recommended drugs?
    - ☐ Yes, always—it's essential for patient safety and informed decision-making
    - ☐ Yes, but only for common or serious side effects
    - ☐ No, this could cause unnecessary anxiety or confusion for users
    - ☐ No, side effect information should only be provided by a doctor or pharmacist

19. How should the kiosk communicate **side effect information** to users? (*Select all that apply*)
    - ☐ Display a list of common side effects on the screen
    - ☐ Provide a printed handout with detailed side effect information
    - ☐ Offer an option to hear side effect information via voice commands
    - ☐ Include a disclaimer advising users to consult a doctor for further clarification

20. In what medical settings do you think an AI-powered drug recommendation kiosk would be **most beneficial**? *(Select all that apply)*
    - ☐ Pharmacies
    - ☐ Emergency Rooms
    - ☐ Clinics & Hospitals
    - ☐ Rural Healthcare Centers
    - ☐ Airports/Transportation Hubs

    Other (please specify): ______

21. What additional features should an AI-powered drug recommendation kiosk include to enhance its **effectiveness**? *(Select all that apply)*
    - ☐ Direct integration with Electronic Health Records (EHRs)
    - ☐ Telemedicine consultation with a licensed doctor
    - ☐ Support for prescription medications (not just OTC drugs)
    - ☐ Real-time alerts for high-risk drug interactions
    - ☐ Multilingual support for diverse populations

    Other (please specify): ______

22. What are the biggest challenges in **scaling AI-powered kiosks** across different healthcare systems? *(Select all that apply)*
    - ☐ High initial investment & maintenance costs
    - ☐ Regulatory approval & compliance challenges
    - ☐ Accuracy & AI model improvements
    - ☐ Data privacy & cybersecurity concerns
    - ☐ Resistance from traditional healthcare providers

### Section 6: AI vs. Traditional Healthcare Services

23. Compared to a **human pharmacist**, how effective do you think an AI-powered kiosk is at providing OTC medication recommendations?
    - ☐ Just as effective as a pharmacist
    - ☐ Nearly as effective, but still needs human verification
    - ☐ Somewhat effective, but lacks personalization
    - ☐ Not effective, humans are irreplaceable

24. Would you trust an AI-powered kiosk more than **online symptom checkers** (e.g., WebMD, Google search, chatbot-based medical apps)?
    - ☐ Yes, AI kiosks are more reliable
    - ☐ Maybe, depending on the technology
    - ☐ No, AI kiosks are just as unreliable as symptom checkers

25. Do you believe AI-powered kiosks will eventually **replace human pharmacists** for OTC medication recommendations?
    - ☐ Yes, AI will replace pharmacists in this role
    - ☐ No, but AI will assist them significantly
    - ☐ No, pharmacists should always be involved

### Section 7: Ethical & Privacy Concerns

26. Do you think AI-powered kiosks should **require user consent** before collecting and analyzing health data?
    - ☐ Yes, explicit user consent should always be required
    - ☐ Yes, but only for sensitive data
    - ☐ No, if the data is anonymized
    - ☐ No, the kiosk should not collect any personal data

27. Do you think AI-powered kiosks should be regulated by **health authorities** (e.g., FDA, WHO) before deployment?
    - ☐ Yes, they should undergo strict regulatory approval
    - ☐ Yes, but only for high-risk medical applications
    - ☐ No, they should be self-regulated by developers

### Final Open-Ended Question

28. What are your biggest concerns or suggestions for improving AI-powered kiosks for medication recommendations? *(Open-ended response)*

**End of Survey**

## ACKNOWLEDGEMENT

We appreciate the invaluable contributions of our healthcare professionals, including Drs. Ahmad Shariftabrizi, Mehdi Maghsudi, Sara Gharibi, Arman Gorgi, Ali Fathi Jouzdani, Amir Mahmoud Ahmadzadeh, Sajad Amiri, Forough Yousefi, Mohammad Zahedi and Nima Sanati for their support and expertise.

## CONFILICT OF INTREST

Sonya Falahati, Morteza Alizadeh, Zhino Safahi, Fatemeh Ghazipour, and Mohammad R. Salmanpour are affiliated with TECVOCO CORP. Company, and the remaining authors have no relevant conflicts of interest to disclose.

## AUTHOR CONTRIBUTIONS

Conceptualization: Sonya Falahati, Morteza Alizadeh and Mohammad R. Salmanpour; Methodology: Sonya Falahati,

Morteza Alizadeh, and Mohammad R. Salmanpour; Investigation: Sonya Falahati, Morteza Alizadeh, and Mohammad R. Salmanpour; Visualization; Sonya Falahati, Morteza Alizadeh, and Zhino Safahi; Resources: Sonya Falahati, Morteza Alizadeh, and Mohammad R. Salmanpour; Validation: Mohammad R. Salmanpour and Navid Khaledian; Writing and editing– original draft, Sonya Falahati, Morteza Alizadeh, and Zhino Safahi; Review: Mohammad R. Salmanpour, Fatemeh Ghazipour, and Navid Khaledian,; Supervision and Lead: Mohammad R. Salmanpour.

## CODE AVAILABILITY STATEMENT

All code developed for this study is publicly available at: https://github.com/MohammadRSalmanpour/AI-powered-Public-Health-Automated-Kiosk-System-/tree/main